# Distance Learning in Primary School During the COVID 19 Pandemic: Results of the "SMART KIDS" Experiment

Svitlana Lytvynova [1], Nataliia Demeshkant [2]

[1] *Institute for Digitalisation of Education of the National Academy of Educational Sciences of Ukraine, 9 M. Berlyns'koho St., Kyiv, 04060, Ukraine*
[2] *Pedagogical University of Krakow, Podchorazych, 2, 30-084, Krakow, Poland*

**Abstract**
The paper analyzes the results of the introduction of the distance learning form (DLF) using electronic educational resources (EER) and the teacher's virtual classroom in primary school. The experiment took place within the framework of the "Smart Kids" All-Ukrainian project during the long quarantine caused by the COVID-19 pandemic. The educational process took place both synchronously and asynchronously. The present paper substantiates the model of organization of distance learning of primary school students using EER and outlines its three main components: the organization of learning; conducting online classes (explaining new material or practicing skills by students) and monitoring the quality of students' independent performance of tasks. The results of the experiment prove that it is necessary to provide teachers and students with computer equipment, Internet access, digital resources for teaching and assessment to implement DLF. It has been established that EER in distance learning can be used both on a regular basis – in each class, and periodically – to explain new material or train skills; the quality of tasks performed by students can be monitored in the virtual office of the teacher and shape an individual trajectory of students' development. The teachers identified the following main problems of DLF implementation: internet interruptions; problems with providing new computer equipment to students and some teachers; lack of state aid in providing EER to all participants in the educational process; limited access to students' computers during complete isolation due to online work of parents. Despite the outlined problems, the quality of distance learning of primary school students during the pandemic using EER was positively and highly assessed by teachers.

## 1. Introduction

During 2020, throughout a prolonged quarantine caused by the COVID 19 pandemic, primary school teachers had significant difficulties in organizing distance communication with students. Firstly, they did not have sufficient skills to apply the distance learning form (DLF), it was a shock for them, and, secondly, despite the fact that students of the XXI century are aborigines of digital technologies, they have not been taught to study by means of distance learning.

In general, research on the implementation of DLF reveals the process of using various digital technologies to training students in higher education. Organizational issues and psychological and pedagogical aspects of the implementation of DLF in general secondary education institutions are being constantly conducted by researchers. Thus, back in 2004, scientists R. M. Bernard, P. C. Abrami, Y. Lou and others conducted a comparative analysis of scientific articles on DLF between 1985 and 2002.

Scientists analyzed more than 232 studies that contained 688 independent results on achievements in the process of DLF implementation; the attitude of participants in the educational process to this form of education; evaluation and preservation of learning outcomes. The authors note that a significant number of DLF programs exceed the results of fulltime learning in the classroom, but there are a significant number of those that do not give significant learning outcomes. The distribution of scientific results by synchronous and asynchronous forms of education showed other results. Asynchronous learning was preferred by students who had significant achievements during full-time study in the classroom, and synchronous – by those who had significant results during DLF. But scientists noted that the overall results showed that the magnitude of the positive effect of the introduction of DLF is essentially zero for all three indicators mentioned above, as there was significant variability in the organization of the learning process, training students and teachers [1].

## 2. Analysis of latest research and publications

With the development of Internet technologies, increasing the overall level of IC-competence of participants in the educational process, areas of research have focused on the study of psychological and pedagogical conditions that impact the effectiveness of learning; study of the components of DLF that affect students' perception of educational material; management aspects, such as decision-making on the use of technology in educational institutions, access to ICT, the ability of users to employ ICT tools in learning [2].

National scholars have made a significant contribution to the study of this problem. Various aspects of DLF are discussed in a number of works, namely by: V. Yu. Bykova (2009), V. M. Kukharenko (2007), K. R. Kolos (2011), V. I. Ko-valchuk (2017), L. M. Lavrynenko (2020), A. F. Manako (2009), N. V. Morze (2010), T. O. Oliynyk (2007), O. M. Pavlenko (2019), S. O. Sysoeva (2009), O. M. Spirina (2007), S. V. Sharova (2019), K. P. Osadcha (2009) at al.

In our opinion, DLF research in general secondary education institutions does not cover all educational aspects – the results of students' learning in DLF are represented by a small number of scientific publications.

Let us note the substantial work, carried out by such researchers as Yu. M. Bohachkov and O. P. Pinchuk (2013) [6]. The authors highlight the problem of building a network of resource centers for distance education to meet the needs of general secondary education institutions; they considered the main tasks of functioning of distance education resource centers and performed a review of possible models of DLF implementation for students of general secondary education institutions.

However, in recent years, the systematic introduction of DLF in the educational practice of general secondary education institutions has not been observed. Educational policies did not encourage general secondary education institutions to introduce distance learning technologies. But in the context of the COVID-19 pandemic DLF has acquired a new meaning in general secondary education institutions and needs additional attention and new research by scientists.

## 3. Problem Statement

In the course of 2020, a number of surveys were conducted concerning the definition of the status of the introduction of DLF, in particular the data of the State Service of Education Quality of Ukraine "Analytical Report on the organization of DLF in general secondary education institutions under quarantine" which analyzed aspects of DLF classes, procedures of getting homework tasks by students and provided recommendations to participants in the educational process (https://cutt.ly/fxm4UEb).

The real state of the problem of using digital tools by teachers during long-term quarantines is substantiated by researchers O. V. Ovcharuk and I. V. Ivanyuk, which is reflected in the work "The state of readiness of general secondary education institutions teachers to use information and educational environment for distance learning during quarantine caused by COVID-19" (https://lib.iitta.gov.ua/719908/) [10].

The monitoring studies of the process of DLF implementation in general secondary education institutions during 2020 identified the following main problems: lack of effective interaction of participants in the educational process; insufficient technical support of students and teachers; inefficient organization of distance learning; weak control of students' academic achievements; poor methodical training of teachers and heads of general secondary education institutions.

The technical and technological component remains the most difficult problem in starting DLF, namely: low speed and quality of Internet connection or its absence; lack of educational digital resources that can fully ensure the formation of students' knowledge of academic subjects; online platform overload; many teachers and students lack modern mobile devices, computers and other gadgets (especially in rural areas); some elementary students cannot work with computers on their own without parental help; limited access of some students to computer equipment was common, too (there may be two to three students and parents working online in a family), etc. [9].

As we can see, the issue of teaching primary school students is a separate item, as the organizational and psychological and pedagogical aspects of teaching primary school students by applying DLF are not fully explored. In addition, during the weakening of quarantine measures period, it was decided that primary school students should study full-time in the classroom in most regions of Ukraine. However, this situation does not solve the problem of organizing DLF for primary school students, and postponing the problem only deepens it.

## 4. Methods of Research

The study presents the results of the introduction of a distance form of education for elementary school students within the framework of realizing the experiment of the All-Ukrainian level named "Smart Kids technology of teaching primary school students", which has been realized in Ukraine since 2017. During 2020, a survey of 94 primary school teachers – participants in the project was conducted on the effectiveness of the use of electronic educational resources during long-term quarantine and distance learning of primary school children. 2454 students and 94 teachers from 12 regions of Ukraine took part in the experiment.

## 5. Research Results

### 5.1. Organization of Distance Learning

In 2017-2020 Ukraine witnessed the implementation of the project "Smart Kids technology of teaching primary school students" (order of the Ministry of Education and Science of Ukraine, 30.08.2017, No.1234), which is based on the method of using electronic educational resources and virtual teacher's office that can fully ensure the effective implementation of DLF in primary school.

Let us consider the model of DLF organization in primary school using electronic educational resources and a virtual teacher's office (Figure 1).

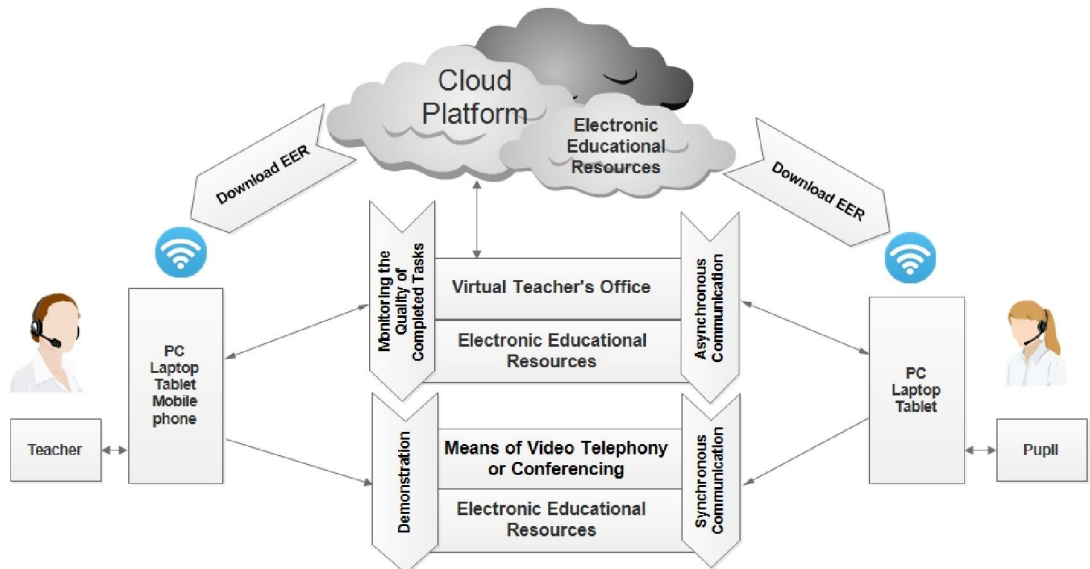

**Figure 1:** Model for the organization of distance learning in primary schools

The introduction of distance learning in primary school is based on the availability of electronic educational resources (EER) for grades 1-4 in Ukrainian language and mathematics of the "Smart Kids" company. Currently, the number of such resources is more than 52 units (http://edugames.rozumniki.ua/catalog/). Teachers and students download electronic educational resources to their gadgets one time at the beginning of the school year. The use of such resources does not require constant access to the Internet.

Distance form of learning, based on the use of EER, includes three main components: the organization of training, conducting online class (explaining new material or practicing skills by students) and monitoring the quality of independent performance of tasks by students. We will further detail all the components.

*Organization of teaching process.*

The main organizational components of distance learning in primary school include the presence of electronic mailboxes of participants in the educational process, means of video telephony or conferencing, electronic educational resources, electronic calendar, access to the Internet (Figure 2).

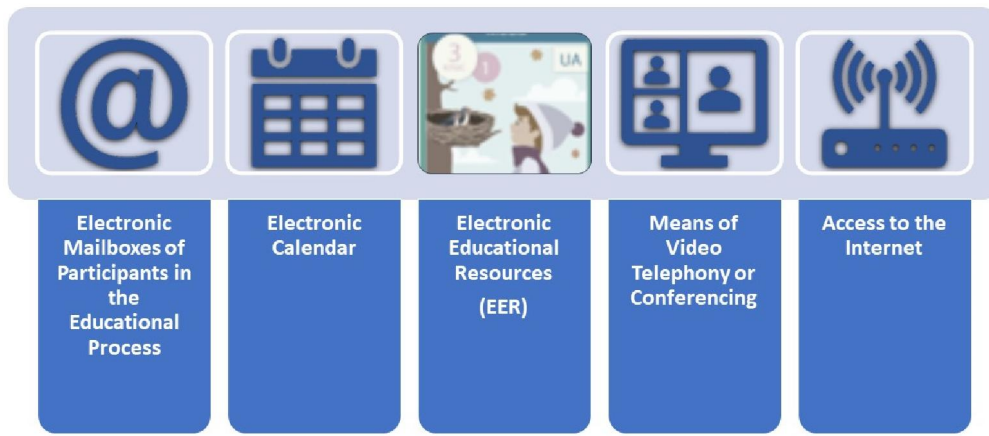

**Figure 2:** Organizational components of distance learning in primary school

All participants in the educational process must be provided with access to the Internet: students – at home, and teachers both at home and in the institution. For online communication, teachers and students need to have the technical means: laptops, personal computers, or tablets. Teachers can use modern mobile phones.

The implementation of online educational communication requires that the teacher be competent to use conference or video telephony, namely: Zoom, Skype, Google Meet, Teams, etc. Training should be provided for students and their parents on the use of online communication tools.

At the beginning of the school year, the teacher needs to form a database of e-mail boxes for students to generate invitations to online classes. Please note that the e-mail box for students in grades 1-4 should be created by parents in order to meet the requirements for the safety of children on the Internet. For example, in the "Age limits on Google accounts" section, Google sets the age limit for creating an email account – 13 years[2].

Organizational aspects include creating a schedule of online meetings with students. Teachers and students need to systematically enter meeting dates into Google Calendar, Outlook Calendar, or Teams Calendar. An effective way is to provide the access of students to the teacher's educational calendar, which will give constant access to current announcements and planned events.

*Online classes.*

According to the requirements of the Sanitary Regulations for schools' online class should last no more than 10 minutes for 1st graders and no more than 15 minutes for 2nd and 4th graders (https://cutt.ly/5cPeWZH). Therefore, a teacher should consider the following organizational aspects of the online classes:

- Rollcall of the students – for this purpose students can write their name in the chat, or the teacher can quickly view the video images of students; the best option – parents inform the teacher about the absence of the child for a good or bad reason, using means of communication, such as a group Viber.
- Explanation of a new material – for this purpose it is necessary to choose in advance those electronic resources which correspond to the object of the class, to think over logical transitions from explanation of a material to demonstration of examples and performance of exercises with students.
- Consolidation of the studied material – the educator should think over the procedure for summarizing the class, organizing feedback with students, announcing homework and the procedure for their assessment (for 3-4 grades).

*Monitoring the quality of student performance.*

Despite the fact that under the Order of the Ministry of Health of Ukraine as of September 25, 2020, No.2205 "On approval of the Sanitary Regulations for general secondary educational institutions" students of grades 1-2 are not recommended to be set mandatory tasks for self-preparation in extracurricular activities, in period of long-term quarantine, the quality of education can be ensured only through a system of homework. An effective way out of this situation is the use of electronic educational resources, which was implemented in the "Smart Kids" All-Ukrainian project.

The quality of the tasks performed by the students was assessed by three indicators: the task was performed correctly, the task was performed with one error, and the task was performed with more than one error. The pupil had the opportunity to redo the task – the motivation was the number of correctly performed tasks (Figure 3).

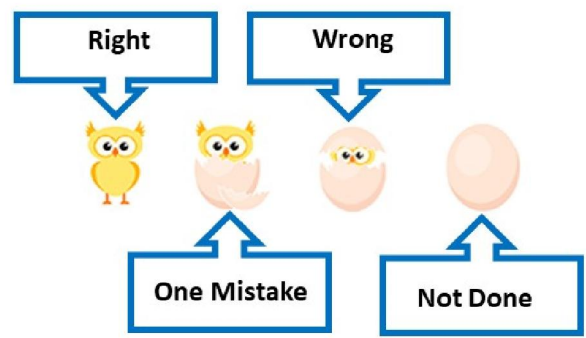

**Figure 3:** Motivational assessment of completed tasks by primary school students

---

[2] https://support.google.com/accounts/answer/1350409?hl=uk

The teacher was able to monitor the quality of the completed tasks in the virtual teacher's office and, if necessary – to assign an additional task to practice the necessary skills (Figure 4).

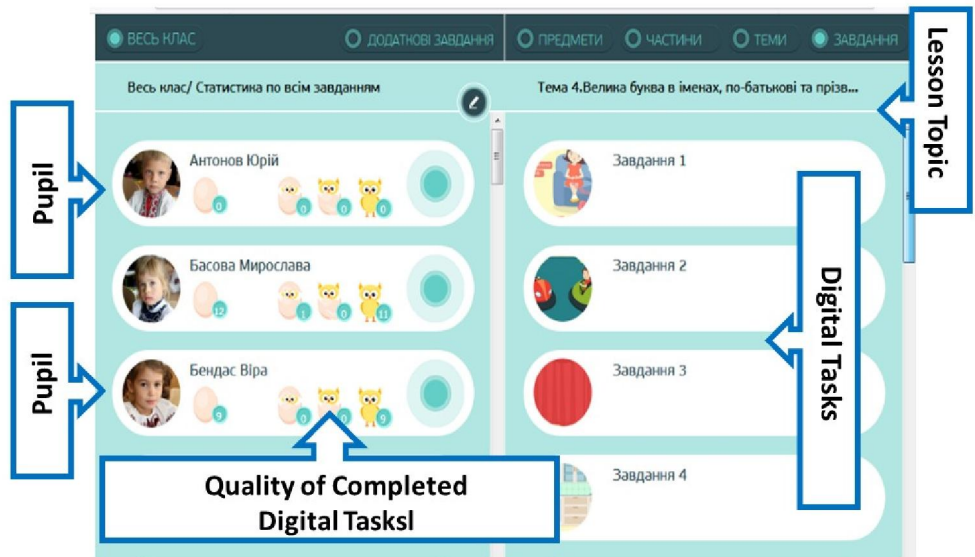

**Figure 4:** Virtual teacher's office – monitoring the quality of tasks performed by students

The procedure for providing an additional task through the virtual teacher's office is quite simple. The task selected by the teacher was marked with a red house, which was displayed on the student's personal computer while working with EER. Thus, while monitoring the students' academic achievements, the teacher was able to shape an individual trajectory of the student's development and provide for the development and consolidation of those tasks that the student encountered difficulties to complete.

### 5.2. Results of the "SMART KIDS" Experiment

At the time of the large-scale COVID-19 pandemic, 94 primary school teachers took part in the experiment to introduce distance education in 12 regions of Ukraine, namely: 1 grade – 27.7%, 2 grade – 22.3%, 3 grade – 17%, 4 grade – 32%. A total of 2454 students took part in the experiment; the distribution of students by grades can be seen in Figure 5.

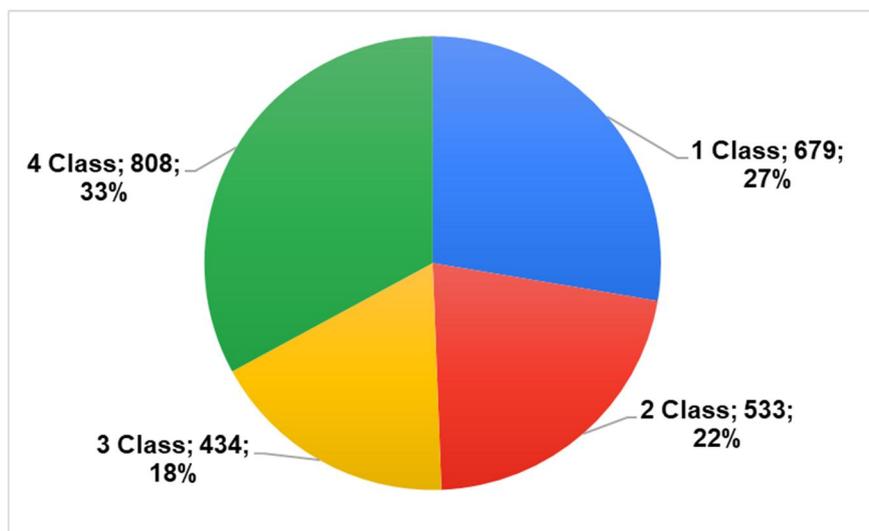

**Figure 5**: Distribution of students by grades with DLF learning process

In the process of organizing the education of primary school students by distance learning, teachers used the following technical means: school and home computers, they also identified a mobile phone and an additional means of instant communication (Figure 6).

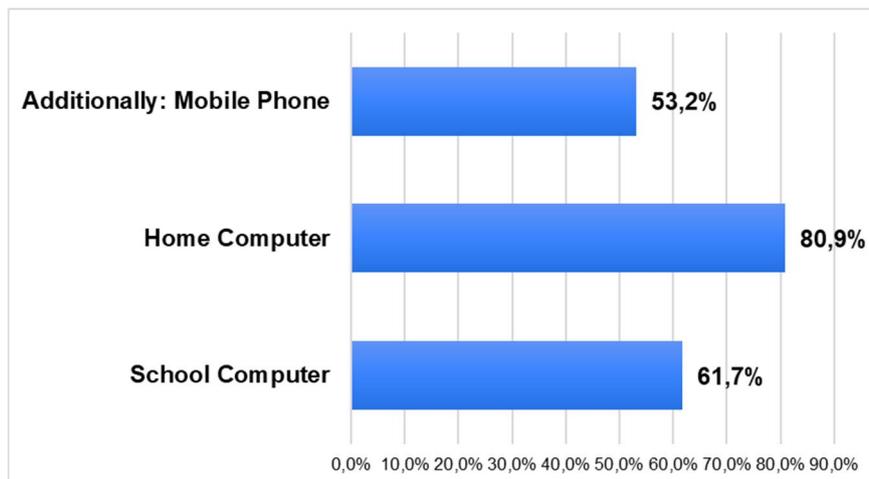

**Figure 6**: Means for organizing distance learning

An important aspect of any form of student learning is assessment. Under the normative documents of the Ministry of Education and Science of Ukraine, assessment in grades 1-2 is not recommended, but teachers – participants of the project carried out constant monitoring of students' academic achievements through a virtual teacher's office [9]. Moreover, they used additional services such as Kahoot, Viber, Microsoft Forms, Learning Apps, and Google Forms to evaluate students in grades 3-4 (Figure 7).

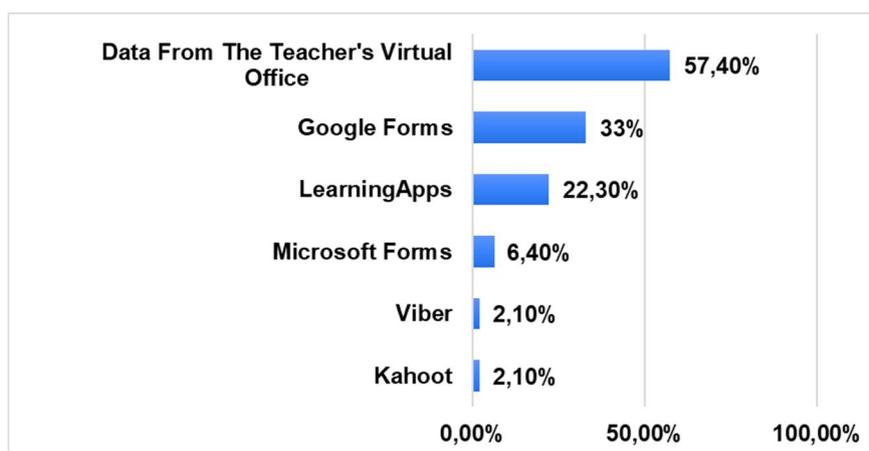

**Figure 7:** Services for assessing the academic achievements of primary school students

As teachers were provided with electronic educational resources, it was important to find out the frequency of using EER during distance learning (Figure 8).

Analyzing the results, we found out that 90.3% of teachers used EER for teaching both continuously and periodically, and only 10% failed to apply EER due to problems with the Internet.

During the pandemic, 64.9% used EER to explain new material, 50% – to test the learning material, 74.5% – gave students additional tasks to practice skills.

To ensure online communication with students, almost 28% of teachers used additional services, such as ZOOM, Google Meet, Teams, and Skype.

The authors also studied the role of EER in teaching primary school students. According to teachers, these were resources for practicing skills – 76.6%, for explaining new material – 64.9%, for monitoring the quality of completed tasks – 50%; 28.8% said that they managed to establish online learning communication and ask the students to speak in front of the class while performing EER tasks.

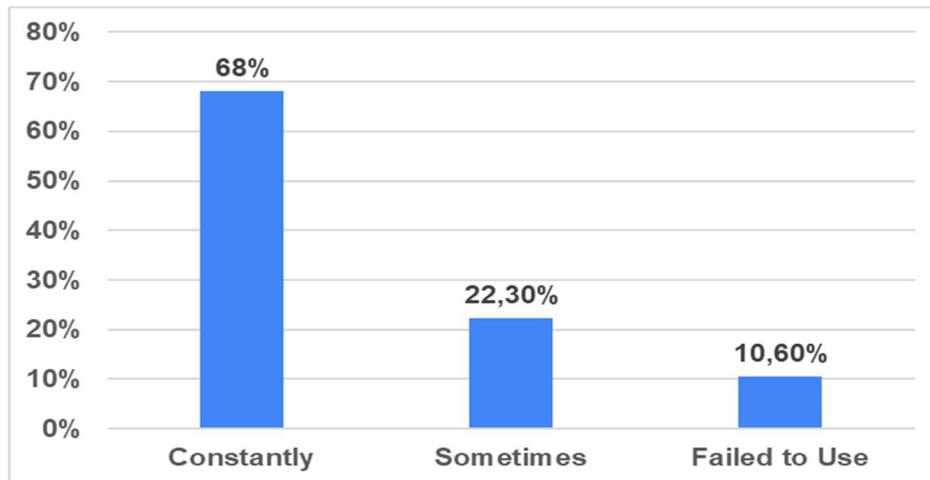

**Figure 8:** Frequency of using Smart Kids EER during the distance learning

Teachers who, in the extreme conditions of transition to distance learning, took the opportunity to use EER to ensure the continuity of student learning, rated the quality of their own work as quite high (Figure 9).

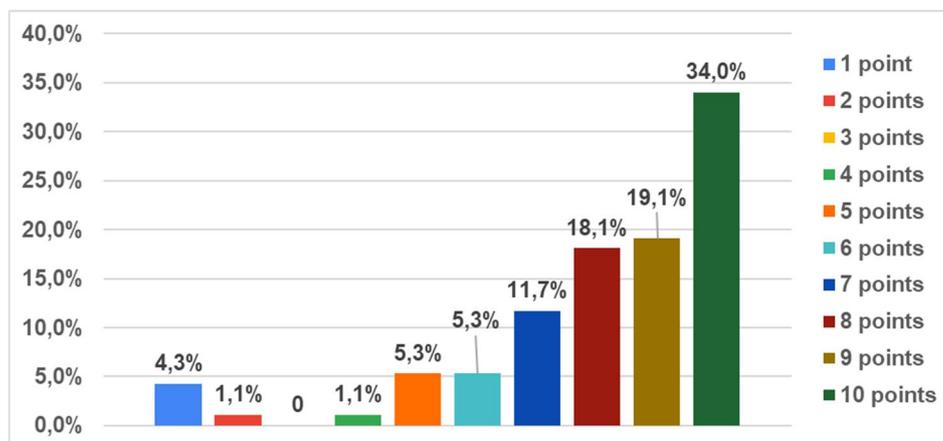

**Figure 9:** Teachers' self-assessment of the use of EER during the distance learning

To evaluate the results of the experiment, it was also important to summarize the teachers' opinions on the need to use EER in primary school in the pandemic event (Figure 10).

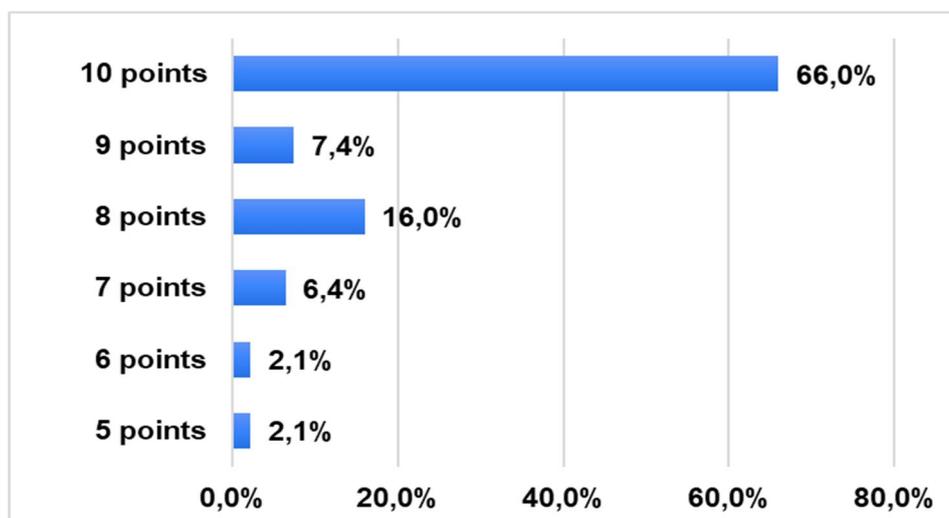

**Figure 10:** Assessing the need to use EER in primary school

Thus, 66% of teachers estimated the need for such resources for primary school at 10 points, 7.4% – at 9 points, 16% – at 8 points and 6.4% – at 7 points. On average, teachers rated the importance of EER at 8.5 out of 10, which is an important indicator in the implementation of Smart Kids technology in educational practice, being an effective means of teaching primary school students.

## 6. Discussion

During the discussion of the situation with COVID-19 and the transition to distance learning, teachers outlined three main problems that hindered the effective use of "Smart Kids" electronic educational resources, namely:
- 27% of teachers acknowledged significant problems with the use of the Internet. They noted unexpected disconnections and disruptions during online classes with students.
- 24% teachers identified the problem of providing computers to students and some teachers. The provision of computer equipment has become prohibitively expensive for low-income families.
- 20% teachers acknowledged the lack of government assistance in providing EER to all students in the classroom.
- Teachers raised the common problem – limited access to students' computers during complete isolation due to parents' online work.

## 7. Conclusions and recommendations for further research

At the moment the primary school system is not only in the process of completing the reform, but also in the process of benchmarking – finding a reference, cost-effective solution to effectively implement distance learning, adopting best practices and implementing best pedagogical practices that will lay the foundations for primary school to provide quality education.

Since EER can be used both in full-time and distance learning, the global experiment on the introduction of distance learning in the pedagogical practice of primary school teachers within the framework of the "Smart Kids technology of teaching primary school students" experiment is a positive example of providing continuous and high-quality student learning.

The developed methods of transition to distance learning in primary school at any time with continuity of education is a significant achievement of cooperation between scientists of the Institute of Information Technologies and Learning Tools of the National Academy of Educational Sciences of Ukraine, "Smart Kids" Holding company and teachers-innovators of primary school in Ukraine.